\documentclass{article}

\usepackage{microtype}
\usepackage{graphicx}
\usepackage{subfigure}
\usepackage{booktabs} 
\usepackage[hyphens]{url}

\usepackage[square,numbers]{natbib}
\usepackage[table,xcdraw]{xcolor}
\usepackage{hyperref}

\usepackage[accepted]{icml2021}

\icmltitlerunning{$\mu$-cf2vec:  Representation Learning for Personalized Algorithm Selection in Recommender Systems}

\begin{document}

\twocolumn[
\icmltitle{$\mu$-cf2vec: Representation Learning for  Personalized Algorithm Selection in Recommender Systems}




\begin{icmlauthorlist}
\icmlauthor{Tomas Sousa-Pereira}{feup}
\icmlauthor{Tiago Cunha}{expedia}
\icmlauthor{Carlos Soares}{feup}
\end{icmlauthorlist}

\icmlaffiliation{feup}{Faculty of Engineering of University of Porto, Portugal}
\icmlaffiliation{expedia}{Expedia Group, Switzerland}

\icmlcorrespondingauthor{Tomas Sousa-Pereira}{t.s.p@fe.up.pt}
\icmlcorrespondingauthor{Tiago Cunha}{tiagodscunha@gmail.com}
\icmlcorrespondingauthor{Carlos Soares}{csoares@fe.up.pt}

\icmlkeywords{Machine Learning, Recommender Systems, Meta Learning, Collaborative Filtering, Representation Learning, Information Retrieval}

\vskip 0.3in
]



\printAffiliationsAndNotice{}  

\begin{abstract}
Collaborative Filtering (CF) has become the standard approach to solve recommendation systems (RS) problems. Collaborative Filtering algorithms try to make predictions about interests of a user by collecting the personal interests from multiple users. There are multiple CF algorithms, each one of them with its own biases. It is the Machine Learning practitioner that has to choose the best algorithm for each task beforehand. In Recommender Systems, different algorithms have different performance for different users within the same dataset. Meta Learning (MtL) has been used to choose the best algorithm for a given problem. 
Meta Learning is usually applied to select algorithms for a whole dataset. Adapting it to select the to the algorithm for a single user in a RS involves several challenges. The most important is the design of the metafeatures which, in typical meta learning, characterize datasets while here, they must characterize a single user. This work presents a new meta-learning based framework named $\mu$-cf2vec to select the best algorithm for each user.  We propose using Representation Learning techniques to extract the metafeatures. Representation Learning tries to extract representations that can be reused in other learning tasks. In this work we also implement the framework using different RL techniques to evaluate which one can be more useful to solve this task. In the meta level, the meta learning model will use the metafeatures to extract knowledge that will be used to predict the best algorithm for each user. We evaluated an implementation of this framework using MovieLens 20M dataset. Our implementation achieved consistent gains in the meta level, however, in the base level we only achieved marginal gains.
\end{abstract}

\section{Introduction}
\label{introduction}
Collaborative Filtering (CF) has become the standard approach to solve recommendations problems \cite{cf:survey}. CF algorithms are relatively simple to implement while having a satisfactory performance in many applications, when compared to other algorithms. Different algorithms have different characteristics, and biases. Their performance varies so choosing the right algorithm is a step towards improving the performance.\\
We have many algorithms that not only perform differently across different datasets, but also even for different users of the same dataset. Discrimination in recommendations is originated from different sources: underlying biases in the input data, or the result of recommendation algorithms \cite{mansoury2019bias}. Different algorithms have different impacts on the recommendations for different groups of users.
The algorithm selection problem has been addressed using Meta Learning (MtL)techniques. The problem is modeled using a set of metafeatures to describe the task and the performance of algorithms using an evaluation measure to describe its behavior. The meta learning model will extract knowledge about the task, and will be used to select the best algorithm. \\
However, the definition of suitable metafeatures is a hard problem. The metafeatures need to be as meaningful as possible to the task. The handcrafted metafeatures efficiency continues to be questioned, since it is difficult to evaluate if they really represent the data.
Representation Learning shares the same goal of the metafeatures, i.e, learn good representation about the data. In Representation Learning we try to extract compact representations from heterogeneous types of data in a numerical representation that can be reused in other learning tasks.\\
We aim to solve the problem of different algorithms that have a different bias in users. To do so, we propose to use Meta Learning. When using Meta Learning we have to extract features about the task we want to solve. We suggest using Representation Learning techniques to retrieve these features. To the best of our knowledge, this solution is the first of its kind: using Representation Learning for Personalized Algorithm Selection in Recommender Systems.\\
This work has two main contributions - 
$\mu$-cf2vec: Representation Learning for Personalized Algorithm Selection in Recommender Systems - which is a framework that uses Representation Learning techniques to retrieve user embeddings, and will be used by a Meta Learning model to solve the algorithm selection problem on the user level. The second contribution is an empirical study of the proposed framework. We implemented a version of the $\mu$-cf2vec framework.

\section{Related Work}
This work uses metalearning techniques to solve the algorithm selection problem on the user level. The metalearning techniques use metafeatures that were retrieved using Representation Learning techniques. To the best of our knowledge, there is no existent research on the problem we aim to solve. 

This work proposes to use Representation Learning to extract metafeatures. This idea has been used in different implementations \cite{DBLP:journals/corr/abs-1809-06120} \cite{jomaa2019dataset2vec}. However, both of these implementations use representation learning on the dataset level.

Our work also proposes to solve the metalearning task of algorithm selection problem on the user level.

There is research in algorithm selection on the user level \cite{DBLP:journals/corr/abs-1805-12118}. This approach uses hand crafted metafeatures to characterize each user. It also uses a set of base learners, each one of them with their own bias. The meta-model will then use the performance label and the handcrafted metafeatures to predict the performance of each algorithm for a new user.

Research has also been done on meta learning at user level using CTR model \cite{luo2020metaselector}. Given a collection of models, the goal is to select the best model for each user. This work uses: model and task agnostic techniques to extract meta features. The meta model will output the best base model for each user by learning the relationships from the metafeatures.

\section{Approach}
\subsection{Problem definition}
Collaborative Filtering is the standard approach to tackle recommendation systems problems. We expect the performance of CF algorithms to vary not only for datasets, but even within the same dataset, for instance, it is different for different users. For each user there is an algorithm that obtains the best performance. The goal is to predict the best algorithm for each user; by doing this we could reduce the biases of the algorithms against groups of users. 

We can use Meta Learning (MtL) to address the algorithm selection problem. From the performances of each algorithm we can use MtL to address different issues: label ranking, selecting algorithm, predicting algorithms performances, selecting hyperparameters and others. For MtL be able to solve the algorithm selection problem on the user-level, we need to characterize the user. This characterization intends to create representation about the users. The representations (metafeatures) will allow the meta models to extract knowledge and recommend the best algorithm for each user.

\subsection{Framework Overview}

\begin{figure}[hbt]
  \centering
    \includegraphics[width=\linewidth]{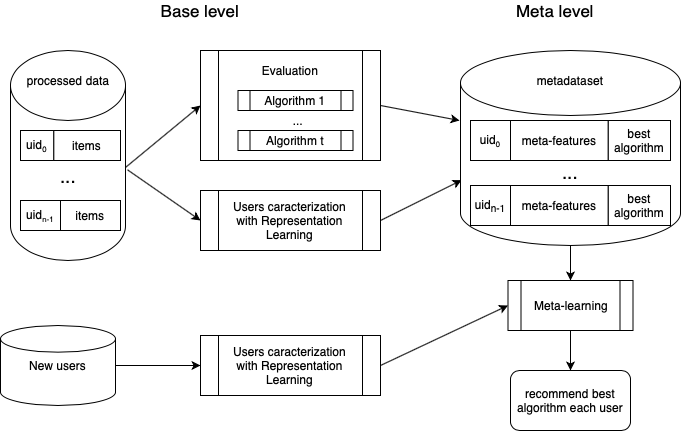}
        \caption{Overview of the $\mu$-cf2vec procedure}
    \label{fig:procedure}
\end{figure}

To try to tackle the problems previously addressed we propose a framework named $\mu$-cf2vec: Representation Learning for Personalized Algorithm Selection in Recommender Systems. This framework proposes the usage of Representation Learning Techniques to automatically retrieve expressive meta features that the meta learning models will use to solve the algorithm selection problem on the user level.
Figure \ref{fig:procedure} presents the procedure this framework follows.

The procedure presented in figure \ref{fig:procedure} have some resembles to the traditional Meta Learning procedure. However, if we analyse carefully we can observe major diferences. While the traditional MtL tasks try to solve the algorithm selection problem on the dataset level, here we propose to solve on the user level. In traditional MtL procedure we would characterize the dataset, and we would recommend the best algorithm for each dataset. In this framework we aim to select the best algorithm for each user. To do so, we have to evaluate the algorithms for each user and characterize the user. The meta learner will then use these information to recommend the best algorithm for each user.

\subsection{Metadata Collection}
Given a CF dataset $d_i$ -- that has a set of users $U$, that have interactions with a set of items $I$ -- we can create a metadataset. Each row of the metadataset relates to an user and it is composed by the metafeatures, and the label. The metafeatures are a set of $k$ values that characterize the user. The metatarget is a label $l_t$ that can be chosen according to the task.

Since this framework proposes to solve a meta learning problem, we need to train and evaluate our models on all users. The splitting strategy is different when comparing with the CF tasks. In this MtL we need to characterize each user, therefore we need interactions from each user. Instead of dividing all the users in three groups: train, validation and test; we divide each user's ratings in train, validation, and test. This splitting is done by the framework.

\subsubsection{Base Level Evaluation}
From a set of base learners $B$, different users have different performances for different base learners. The base learners are trained using the training interactions from all users. The validation interactions are used to tune the parameters and the test interactions are used to make an unbiased evaluation of the model. There is an evaluation of every base learner for each user.

The evaluation is linked to the metatarget of each user. This evaluation is chosen accordingly to the task we want to solve. The metatarget represents the meta-variable that the meta learner wishes to understand or predict i.e. the algorithm with the best performance for a given task. We can select the metatarget to rank the base learners, select the best base learner, predict the performance of each base learner and others. This makes the framework flexible. We can select the meta target to optimize the task we want to solve. 

\subsubsection{Retrieving metafeatures}

One of the contributions of this framework is using Representation Learning to retrieve metafeatures at the user level. The RL techniques allow us to automatically generate metafeatures for each user. Using this approach we remove the need of manually creating metafeatures. Therefore, $\mu$-cf2vec proposes to use Representation Learning techniques to retrieve the metafeatures for each user.   

Using these techniques to automatically retrieve metafeatures for each user, is firstly introduced in this framework. $\mu$-cf2vec creates representations for all users using RL techniques. These representations are created using the train interactions from all users. The metafeatures created by applying these RL techniques are going to be used by the meta model to select the best algorithm for each user.

\subsubsection{Inference}

In real world systems, this framework has to be trained before its deployment. The training will be used to select the best meta learners for the task. During the training process the representation learning technique is chosen. The evaluations of the multiple base learners are also obtained in this stage. The training stage will create the metadataset and select the best meta model.

After training the framework, it can be deployed and will automatically update itself. The framework will retrieve the user embeddings periodically, using the newly collected user-item interactions. These embeddings will be retrieved using the best RL technique found on the training step. The meta learner will then extract knowledge which will provide the ability to make personalized algorithm recommendations. For the new users or existing users, for which the system is not able to compute embeddings, the framework will use a naive strategy by recommending the Most Popular algorithm.

\section{Experimental setup}
This sections presents the implementation of the previously proposed framework.

\subsection{Base Level}
\subsubsection{Base learners}
We used the Fast Python Collaborative Filtering Framework \cite{implicit} from which we are going to use the following algorithms:
\begin{itemize}
  \item Alternating Least Squares (ALS)
  \item Bayesian Personalized Ranking (BPR)
  \item Logistic Matrix Factorization (LMF)
   \item Most popular (to be used as baseline)
\end{itemize}

Using these algorithms we can capture different biases. We have model-based and memory-based Collaborative Filtering techniques. The first three base learners are matrix factorization techniques. They have a parameter that is the number of latent factors to be computed by each model. Once we decided on the splitting strategy and the base learners we would use, we then needed to understand the performance of the base learners to decide upon the split data percentages.

\subsubsection{Dataset}
In this implementation we decided to use the MovieLens 20m to tackle this problem. We transformed it into an implicit dataset by setting a threshold for ratings greater than 3.5. For this implementation we are not going to consider the Cold Start Problem. For this reason we will remove all users and items with less than 10 interactions. To evaluate the base learners we decided to user NDCG@K.

There are multiple different strategies to split the data. This is a metalearning task, hence we want the meta model to be trained on all users. The meta learning tasksrequire all the users to train the model. Each users' ratings are divided in the three groups: train, validation and test. In meta learning tasks it is relevant to define the different percentages of each group. These are selected according to the problem we want to solve. In the meta level we only use the training ratings. The other two groups are only used in the base level.

We have 4 different algorithms and we still may face the problem that in some users none of the algorithms will have a good performance. There may exist a group of users, for which all of the base learners have a NDCG of 0. We are going to call this class of users \emph{zeroes}. To select the best data split we retrieved the NDCG of all models for each user. We selected $K$ = 30, 70\% training, 10\% validation and 20\% for the testing. These values were selected by finding a balance between the number of users belonging to class \emph{zeroes} and the value of $K$. Since choosing a high value of $K$ reduces the number of \emph{zeroes} users, and the \emph{zeroes} users can introduce errors in our models.

\subsubsection{Obtaining User-level metafeatures based on embeddings }

In this work we are going to use state of the art techniques in Representation Learning:  Denoising Auto-Encoders for Top-N Recommender Systems  \cite{denoising} and Variational Autoencoders for Collaborative Filtering \cite{liang2018variational}. 

For the Variational Auto-encoders (VAE) we are going to use the code from \cite{nvidia:vae}. Since this  model is designed for Collaborative Filtering it splits all the users in Train (80\%), Validation (10\%) and Test (10\%). In our approach, as it is a Meta Learning task, we use all the users in every step and in each user we split the user ratings in  Train (70\%), Validation (10\%) and Test (20\%). The splitting strategy of the \cite{nvidia:vae} is integrated with the model, however, we need to obtain the metafeatures for all users. To retrieve the embeddings we fed the model \cite{nvidia:vae} with all of our training data. We let the model train and split the data in its own partitions. After the model was trained we obtained the embeddings for all the users. The model creates a representation of size 200 for each user. All the users training interaction enter the VAE. These users are automatically splitted in 80\% for training, 10\% for validation and 10\% for test. The training users are used to train the models and the parameters are tuned with the validation users. After the model is trained we retrieve the embeddings for all the users. 

For the Collaborative Denoising Auto-Encoders (CDAE) we are going to use the DRECPY \cite{drecpy} library. In the original paper of this model the author facilitated the retrieval of the factors for each user. The attribute \(V\) has the user embeddings. This attribute has a dimension of \( U x K\), U being the number of users and K the hidden factors. We are going to use two different hidden factors: 50 because it is the number of hidden factors used by the Author, and 200 hidden factors. We retrieve the embeddings of size 200  to make a fair comparison with the Variational Auto-encoders which have size 200. 

We are going to have three metadatasets: VAE, CDAE50, and CDAE200. Each one of them is built using the metafeatures obtained by RL techniques with the same name, e.g., the VAE dataset is built using metafeatures retrieved with the Variational Auto-Encoder and with the metatarget that is the name of the base learner with the best performance.

\subsection{Meta-level Experimental Setup}

These metadatasets are going to be used to solve the meta-level problem. We are going to build meta models which will leverage these metadatasets to select the best algorithm for each user. To achieve this, we are going to implement and evaluate a wide variety of models. We will then select the most suitable for our problem. The meta learners we are going to use are the following:

\begin{itemize}
  \item Logistic Regression \cite{defazio2014saga}
  \item Multi Layer Perceptron\cite{HINTON1989185}
  \item Support Vector Machine  \cite{Platt99probabilisticoutputs}
  \item Random Forest \cite{10.1023/A:1010933404324}
  \item Light GBM \cite{NIPS2017_6907}
\end{itemize}

These algorithms have different characteristics and different biases. We also tried to use Knn however due to the characteristics of the data we couldn't achieve good results. We performed a grid search to tune the hyper parameters of the meta models. We also retrieved the results for normalized data using z-score normalization. We used Cross Validation with 5 Folds, since this technique allows a deeper look into the model performance. Comparing with the traditional  train-test split, this approach reduces the variance. Together with the cross validation we used a different variety of metrics: accuracy, precision, recall and all their averages. These metrics are acquired in the meta level and provide analysis for the performance of the meta models in each class. To evaluate the performance of the meta models in the base level, we are going to understand the impact of the meta models' predictions in the base level performance. The impact in base level performance will be in terms of NDCG and the intuition behind it is the following: assume for user $u_i$ ALS has a NDCG of 0.4, BPR a NDCG of 0.35, and Most popular a NDCG 0.2. The best algorithm might be the ALS but if our meta model selects BPR, the overall performance for that user would be similar to the correct label ALS. Nevertheless, if the meta model selects Most Popular, the performance on that user will be affected.

\section{Research Questions}

\subsection{RQ1: Can we use Representation Learning to obtain useful metafeatures on the User Level?}
\label{rq1}
This project addresses several new challenges. It is the first, to the best of our knowledge, that uses meta learning together with Representation Learning in order to solve the algorithm selection problem on user level. Using Representation Learning techniques to automatically retrieve metafeatures for each user,  was firstly introduced in this work. This RQ will answer the question whether it is possible to retrieve expressive metafeatures about the users.

\begin{figure}[hbt]
  \centering
    \includegraphics[width=\linewidth]{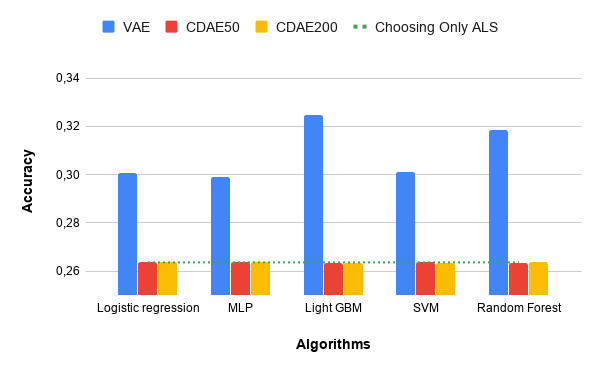}
        \caption{Meta Level Accuracy score on normalized datasets}
    \label{fig:accuracy-norm}
\end{figure}

We know that we have five classes, so we can assume that a completely random model would have an accuracy of 20\%. Table \ref{tab:Proportions_class} presents the distribution of the classes, where ALS is the biggest with a value of 26,37\%. For our model to have an impact on solving this problem, we have to be better than a random model (that has an accuracy of 20\%) and have to improve the result of only recommending one class (better than 26,37\%). $\mu$-cf2vec proposes to remove the biases presented in CF algorithms, hence the meta models have to predict more than one class. Figure \ref{fig:accuracy-norm}, which presents the accuracy for the models with normalization, can help us answer this question.

\begin{table}[!htb]
\centering
\begin{tabular}{|l|ll|}
\hline
\rowcolor[HTML]{EFEFEF} 
\textbf{Algorithms} & \textbf{\# users} & \textbf{Percentage} \\ \hline
ALS & 34215 & 26,37 \\
Most Popular & 29088 & 22,42 \\
BPR & 25925 & 19,98 \\
Zeroes & 21277 & 16,40 \\
LMF & 19252 & 14,84 \\
Total & 129757 & 100 \\ \hline
\end{tabular}%
\label{tab:Proportions_class}
\caption{Proportions of each class}
\end{table}

In the CDAE datasets the models achieved an accuracy lower or equal to 26,37\%. Table \ref{tab:logistic-confusion} presents the confusion matrix for the Logistic Regression model for the normalized CDAE50 dataset. The confusion matrix indicates that the model only recommends one class, \emph{ALS}. The recall of this class is 1, while in the other classes it is around 0. The accuracy of the models using both CDAE metadatasets is the same as the percentage that the ALS class presents in table \ref{tab:Proportions_class}. The results of the other meta models are similar.

\begin{table}[!htbp]
\centering
\begin{tabular}{|
>{\columncolor[HTML]{EFEFEF}}l |l|l|l|l|l|}
\hline
 & \cellcolor[HTML]{EFEFEF}\textbf{ALS} & \cellcolor[HTML]{EFEFEF}\textbf{BPR} & \cellcolor[HTML]{EFEFEF}\textbf{LMF} & \cellcolor[HTML]{EFEFEF}\textbf{MostPopular} & \cellcolor[HTML]{EFEFEF}\textbf{Zeroes} \\ \hline
\textbf{ALS} & 6837 & 0 & 0 & 4 & 2 \\ \hline
\textbf{BPR} & 5182 & 0 & 0 & 1 & 1 \\ \hline
\textbf{LMF} & 3847 & 0 & 0 & 2 & 1 \\ \hline
\textbf{MostPopular} & 5812 & 0 & 0 & 3 & 3 \\ \hline
\textbf{Zeroes} & 4249 & 0 & 0 & 3 & 4 \\ \hline
\end{tabular}
\caption{Logistic Regression meta model confusion matrix for the normalized CDAE50 dataset.}
\label{tab:logistic-confusion}
\end{table}

When using the models on the VAE metadataset, every meta model improves the accuracy score of 26,37\%. The best model is the Light GBM with normalized data, achieving an accuracy of 32,48\%. When analyzing the results in table \ref{tab:class-report-lgb}  we can identify some  aspects. The LMF class is the one with the worst overall score, which can be due to the low proportion of the class as stated in the table \ref{tab:Proportions_class}. This model has a best overall score in the class \emph{zeroes}, even though it is the fourth in amount of appearances. The precision score and the F1 score for this class are the best in comparison with the other classes. In the metric recall it is in second place, just behind the most common class \emph{ALS}. These results indicate that the \emph{zeroes} class affect the performance of the model. Since ideally we would want to recommend different algorithms, i.e. it would be better if the the meta models have better performance in other classes instead of the \emph{zeroes} class. There is another model in this same dataset that has almost the same accuracy but its recommendations spectrum is different. 

\begin{table}[!htb]
\centering
\resizebox{\linewidth}{!}{
\begin{tabular}{|
>{\columncolor[HTML]{EFEFEF}}l |l|l|l|}
\hline
\textbf{} & \cellcolor[HTML]{EFEFEF}\textbf{Precision} & \cellcolor[HTML]{EFEFEF}\textbf{Recall} & \cellcolor[HTML]{EFEFEF}\textbf{F1} \\ \hline
\textbf{ALS} & 0.31 & 0.55 & 0.40 \\ \hline
\textbf{BPR} & 0.31 & 0.21 & 0.25 \\ \hline
\textbf{LMF} & 0.23 & 0.02 & 0.04 \\ \hline
\textbf{MostPopular} & 0.32 & 0.31 & 0.31 \\ \hline
\textbf{Zeroes} & 0.40 & 0.40 & 0.40 \\ \hline
\multicolumn{1}{|c|}{\cellcolor[HTML]{EFEFEF}\textbf{--}} & \multicolumn{1}{c|}{\textbf{--}} & \multicolumn{1}{c|}{\textbf{--}} & \multicolumn{1}{c|}{\textbf{--}} \\ \hline
\textbf{Macro avg} & 0.31 & 0.30 & 0.28 \\ \hline
\textbf{Weighted avg} & 0.32 & 0.32 & 0.30 \\ \hline
\end{tabular}
}
\caption{LightGBM classification report on normalized VAE}
\label{tab:class-report-lgb}
\end{table}

The Random Forest meta model has an accuracy of 31,85\%. While analyzing the results in the table \ref{tab:class-report-rf} we have a new scenario in comparison with the previous meta model. Random Forest model has a better score when predicting the class \emph{ALS}. The metrics F1 and Recall on the class \emph{ALS} are the overall best. The overall performance of this model has improved in comparison with Light GBM.

\begin{table}[!htb]
\centering
\resizebox{\linewidth}{!}{
\begin{tabular}{|
>{\columncolor[HTML]{EFEFEF}}l |l|l|l|}
\hline
\textbf{} & \cellcolor[HTML]{EFEFEF}\textbf{Precision} & \cellcolor[HTML]{EFEFEF}\textbf{Recall} & \cellcolor[HTML]{EFEFEF}\textbf{F1} \\ \hline
\textbf{ALS} & 0.30 & 0.75 & 0.42 \\ \hline
\textbf{BPR} & 0.36 & 0.13 & 0.19 \\ \hline
\textbf{LMF} & 0.40 & 0.0014 & 0.0028 \\ \hline
\textbf{MostPopular} & 0.34 & 0.21 & 0.26 \\ \hline
\textbf{Zeroes} & 0.40 & 0.30 & 0.34 \\ \hline
\multicolumn{1}{|c|}{\cellcolor[HTML]{EFEFEF}\textbf{--}} & \multicolumn{1}{c|}{\textbf{--}} & \multicolumn{1}{c|}{\textbf{--}} & \multicolumn{1}{c|}{\textbf{--}} \\ \hline
\textbf{Macro avg} & 0.36 & 0.28 & 0.24 \\ \hline
\textbf{Weighted avg} & 0.35 & 0.32 & 0.26 \\ \hline
\end{tabular}%
}
\caption{Random Forest classification report on normalized VAE}
\label{tab:class-report-rf}
\end{table}

\begin{figure}[hbt]
  \centering
    \includegraphics[width=\linewidth]{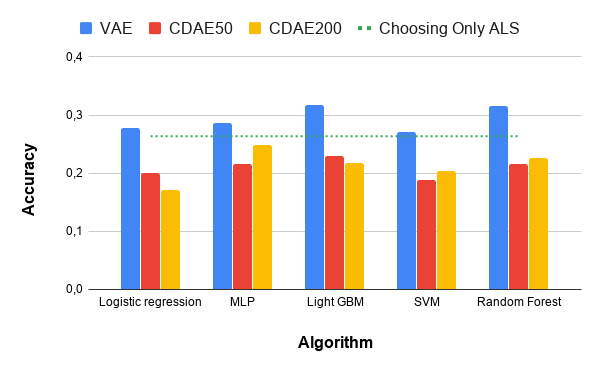}
        \caption{Meta Level Accuracy score on normalized datasets}
    \label{fig:accuracy-norm-smote}
\end{figure}

In table \ref{tab:Proportions_class} we can observe that the classes are not evenly distributed. To overcome this problem, we used the SMOTE technique to create synthetic data. Although we expected to achieve better results with the use of this technique, the results presented in figure \ref{fig:accuracy-norm-smote} do not indicate that. In figure figure \ref{fig:accuracy-norm-smote}  the performance is consistently worst when compared to the figure \ref{fig:accuracy-norm}.

Based on these results we can deduce that on the meta level we can achieve an improvement in the performance. It is possible to use Representation Learning techniques to make personalized algorithm selection in recommender systems. 
From these results we can infer some points: the best overall meta model is the Light GBM (it has the best accuracy and the best performance for the \emph{zeroes} class), although the Random Forest has its own advantages (it has a good performance for the \emph{ALS} class). Through Light GBM, we can infer that, although class \emph{zeroes} is the fourth most common class, it has the best performance for this model. That shows how this class can mislead the results. 

\subsection{RQ2: Which Representation Learning Technique is the best?}
As stated previously this project introduces a completely new approach to the algorithm selection problem on the user level. We proposed to use meta learning combined with  Representation Learning, in order to retrieve meta features. Later, these meta features will be used by the meta model to try to predict the best algorithm for each user. 
For this project we have chosen two techniques: Variational Auto-encoders and  Denoising Auto-encoders.

From these two techniques we have three experimental setups as stated previously. The VAE metadaset used the user embeddings retrieved with the Variational Auto-encoder and has a size of 200. The CDAE dataset is created with users representations acquired by the Denoising Auto-encoders. We have two different datasets from these techniques: CDAE50 and CDAE200. The original paper of this implementation found out that the optimal number of factors is 50, therefore we started by obtaining the results with 50 factors. However, the VAE dataset is composed by representations with size 200. To have an head-to-head comparison between the two techniques we retrieved the CDAE representations with size 200.  

\begin{figure}[hbt]
  \centering
    \includegraphics[width=\linewidth]{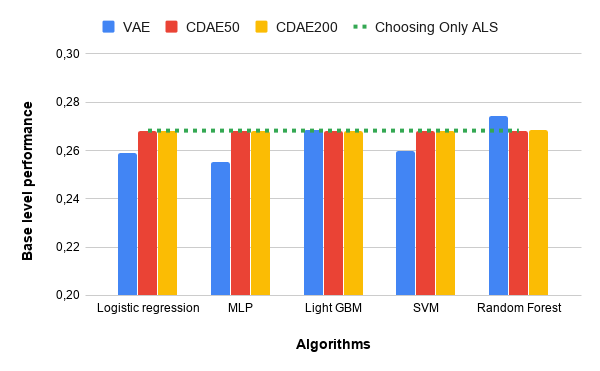}
        \caption{Base Level Performance on normalized datasets}
    \label{fig:base-level-norm}
\end{figure}

To identify which of the techniques is more suitable to solve this problem, we have to understand what their performance is. We have different techniques as stated previously: VAE and CDAE. From the figures presented before (\ref{fig:accuracy-norm} and \ref{fig:base-level-norm}) the VAE metadataset achieves better results in every single meta model. These results are consistent in all metadatasets: normalized or non normalized, with or without zeroes and with or without the SMOTE technique. The results presented indicate that the Variational Auto-encoder is the better implementation in this situation.

\subsection{RQ3: What is the impact on the base level performance achieved by $\mu$-cf2vec?}
$\mu$-cf2vec is a framework which aims to improve the recommendations' performance of every user. As stated previously, the meta  is able to predict the best algorithm for each user with an accuracy higher than the baselines. However, we need to look deeper into the performance affected by the users, in order to understand whether the model can improve the performance for the users or not.

In table \ref{tab:ndcg_eval} we have the upper and lower bounds for our model.  The \emph{Perfect NDCG} is the upper bound of our model. This value is the average of NDCG of the best algorithm (with higher NDCG) for each user. The other rows represent the value of the NDCG of a meta model that has only selected one single algorithm (ALS, BPR, etc). One thing we have to revisit is that the class \emph{zeroes} has zero NDCG in every algorithm. As stated previously $\mu$-cf2vec will recommend the Most Popular algorithm to users identified as \emph{zeroes}. This is the policy of the framework to address users that we don't have enough information to make good recommendations. Thus, the goal of our meta model is to improve the value of 0,27 presented in table \ref{tab:ndcg_eval}. If we improve this score, we can infer that the performance is better than if we only used the ALS. 

\begin{table}[!htb]
\resizebox{\linewidth}{!}{%
\begin{tabular}{|l|l|}
\hline
\rowcolor[HTML]{EFEFEF} 
\textbf{Algorithms} & \textbf{NDCG Evaluation} \\ \hline
Perfect NDCG & 0,46 \\
choosing only ALS & 0,27 \\
choosing only BPR & 0,21 \\
choosing only LMF & 0,17 \\
choosing only Most Popular & 0,21 \\ \hline
\end{tabular}%
}
\caption{Base level performance in terms of NDCG. }
\label{tab:ndcg_eval}
\end{table}

From figure \ref{fig:base-level-norm} we can observe some interesting aspects. At the meta level the VAE dataset was always the best. However, at the base level, only Light GBM and Random Forest achieved better results with VAE. For the other algorithms the CDAE datasets achieved a better perfromance when comparing to the VAE performance.

When using  the VAE dataset, which can be more expressive as stated previously, the meta models have a wide spectrum of recommendations. While in the meta level we could have improvements by using the VAE and recommending different algorithms, in the base level it can be worse. For instance, the MLP meta model has an accuracy of around 30\% using the VAE dataset compared to an accuracy of around 26\% for both CDAE datasets. From this result we could expect that in the base level the performance would also be better when using the VAE dataset. However, the figure \ref{fig:base-level-norm} indicates the opposite: the base level performance in MLP using the VAE is worse than both CDAE. While using the VAE dataset the MLP is recommending different algorithms, in the CDAE datasets it is only recommending \emph{ALS}. In this case it achieves a better performance. 

When inspecting figure \ref{fig:base-level-norm}, we can observe two models where the base level performance is better than if we only used \emph{ALS}. The Random Forest and the Light GBM models both achieve marginal gains over recommending the best single class. 

From these results we only achieved marginal gains in the base level performance. Even though the VAE dataset achieved the best meta level performance, this was not true on the base level. In fact, in the base level (in some cases) it is better to use the CDAE datasets. Sometimes it is better to select one single algorithm to achieve the best performance, than selecting multiple algorithms.

\section{RQ4: What is the impact of the zeroes class on $\mu$-cf2vec?}
As expected, there are some users for which none of the algorithms considered provide any useful recommendation. These users belong to the class \emph{zeroes}, since the maximum NDCG that all base learners algorithms' could achieve was zero.The users with this class are a group of users that we do not have enough information on to solve their problem. This can be related to the Cold Start Problem.

In research question \ref{rq1} we presented that even though the \emph{zeroes'} class is in fourth place, in terms of occurrences, most of the models could detect this class more easily than the BPR and Most Popular (although both are more frequent). These results might indicate that these users are misleading the models performance. $\mu$-cf2vec aims to recommend the best algorithm for each user. Since the \emph{zeroes} users do not have good performance in any algorithm, we decided to remove them. This will allow us to understand their impact on the meta models performance.     Table \ref{tab:Proportions_classnozereos} presents the proportions of data after we removed all the users that belong to the class of \emph{zeroes}. This analysis was made to understand the impact of the class \emph{zeroes}. 

\begin{table}[!htb]
\centering
\begin{tabular}{|l|ll|}
\hline
\rowcolor[HTML]{EFEFEF} 
\textbf{Algorithms} & \textbf{\# users} & \textbf{Percentage} \\ \hline
ALS & 34215 & 31,54 \\
Most Popular & 29088 & 26,81 \\
BPR & 25925 & 23,90 \\
LMF & 19252 & 17,75 \\
Total & 108480 & 100 \\ \hline
\end{tabular}%
\caption{Proportions of each class}
\label{tab:Proportions_classnozereos}
\end{table}

\begin{figure}[hbt]
  \centering
    \includegraphics[width=\linewidth]{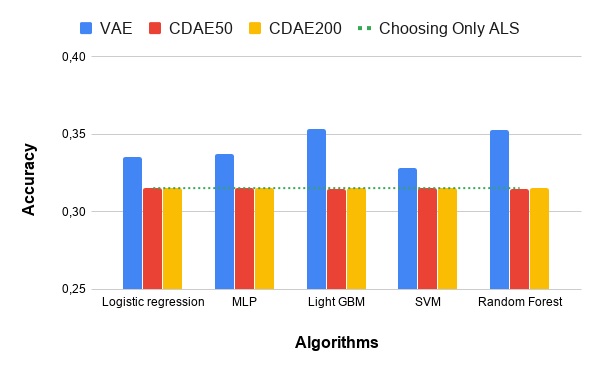}
        \caption{Accuracy score on normalized datasets, without zeroes}
    \label{fig:accuracy-norm-nzers}
\end{figure}

\begin{figure}[hbt]
  \centering
    \includegraphics[width=\linewidth]{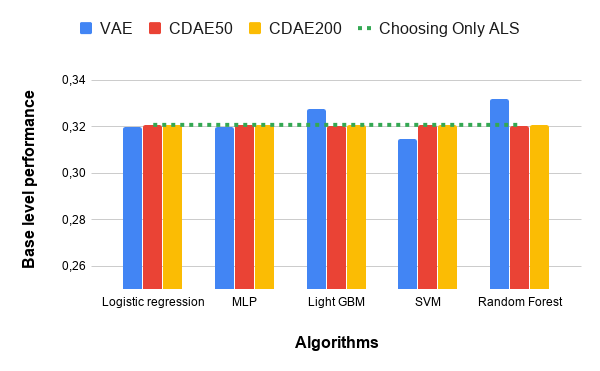}
        \caption{Base Level Performance onnormalized datasets, without zeroes}
    \label{fig:base-level-norm-nzers}
\end{figure}

Figures \ref{fig:accuracy-norm-nzers} and \ref{fig:base-level-norm-nzers} present the accuracy and base level performance for each algorithm in all metadatasets without users belonging to class \emph{zeroes}. Since we removed the users labeled as \emph{zeroes}, the distribution of the classes has changed, as we can see in table \ref{tab:Proportions_classnozereos}. When comparing to the distribution with all users (table \ref{tab:Proportions_class}) we can highlight some points. The perfect meta model would increase the NDCG evaluation by 20\%. The base level lower bound is the most common class and it still is \emph{ALS} with a percentage of 31,54\%. The overall improvement in the base level performance was around 20\%. These values by themselves show the impact of the users in the overall metrics. These are the reference results we have to use when analyzing the ones without the users labeled as \emph{zeroes}.

The meta level accuracy scores are similar to the results with all users. The VAE dataset still has the best overall performance when comparing to the CDAE. The best models are the Light GBM and the Random Forest achieving an accuracy of around 35\%. This accuracy showed some improvements, however, we have to recall that the new percentage of the most common algorithm \emph{ALS} is 31,54\%. From these results the improvement is not significant. For the CDAE datasets the results are similar, and the meta models only recommend the most common class, achieving an accuracy of around 31,54\%.

The base level performance results in figure \ref{fig:base-level-norm-nzers} are similar to the results with all users (shown by figure \ref{fig:base-level-norm}). The only two models that improved the base level performance are the Random Forest and the Light GBM, when comparing to the ones that only used the most common class. These marginal gains where achieved by using VAE dataset, which reinforces the hypothesis that the VAE technique is able to achieve better results. One aspect where the results in figure \ref{fig:base-level-norm-nzers} differ from the ones in figure \ref{fig:base-level-norm} is the fact that the Logistic Regression and the MLP are closer to the lower boundary. This is explained by the fact that False \emph{zeroes} users identified before are now being identified as \emph{ALS}. The CDAE dataset has similar results when using all users. All the models are recommending the \emph{ALS} class, hence the base level performance will be around the same as the \emph{ALS} class.

We can conclude, from these results and the ones in the research question \ref{rq1}, that this organically created class can bring problems to this framework. There are different ways to try to minimize the effects of this class. They will be discussed in the future work.

\section{Conclusion}

$\mu$-cf2vec proposes to solve the algorithm selection on the user level. To do so, it uses a Meta Learning approach by learning from prior experience. The meta learners leveraged from a dataset which is built with the users metafeatures and the user label. The metafeatures are automatically retrieved by using Representation Learning techniques. Using these techniques we removed the need of manually selecting the metafeatures, a process that can be slow and inefficient. The meta target is related to the algorithms we want to recommend. The meta models use the metafeatures to recommend the best algorithm for each user.

This implementation of the framework generated metafeatures from two Representation Learning techniques: Variational Auto-encoders and Denoising Auto-encoders.  In this implementation the metatarget is obtained by selecting the name of the algorithm that achieved the best performance for each user. We used a set of 5 meta models to try to solve the algorithm selection problem on the user level. In every meta model the dataset with metafeatures obtained using Variational Auto-encoders achieved the best meta level results. From this study we can conclude that we are able to achieve meta level improvements by using Representation Learning (using VAE). The meta models Light GBM and Random Forest are the ones with the best overall performance. In the base level we only achieved marginal gains. The organically created \emph{zeroes} class has a negative impact on the meta models performance. We can relate this class to the cold start problem. This implementation is the first of its kind when addressing Personalized Algorithm Selection in Recommender Systems.  

\subsection{Limitations and Future Work}

$\mu$-cf2vec proposes an innovative approach to solve Personalized Algorithm Selection in Recommender Systems. This implementation of the framework has some limitations that can lead to future work:

\begin{itemize}
    \item In the implementation of the $\mu$-cf2vec framework we only used a single dataset. To validate this framework, we have to test it in more datasets.
    \item We only used two Representation Learning techniques. However, there are many different techniques. In future work we should use different RL techniques.
    \item  We did a grid search to find the best parameters, however there is a chance that the hyperparameters we have chosen aren't the best for the task we aim to solve. Not having the optimal parameters for a meta model can have a negative impact on the performance.
    \item The base learners selected are a limitation of this work. The framework $\mu$-cf2vec wants to remove the bias of the CF algorithms. We have to choose a variety of base learners to leverage different biases. To address this limitation we should have a more diverse group of base learners. 
\end{itemize}

\bibliography{refs}

\end{document}